 \documentclass[aps,prr,reprint,twocolumn,superscriptaddress,floatfix,nofootinbib,amsmath, amssymb, longbibliography]{revtex4-1}
\usepackage[utf8]{inputenc}
\usepackage{graphicx}
\usepackage{dcolumn}% Align table columns on decimal point
\usepackage{bm}
\usepackage{color}

\begin{document}
\title{Polymer Physics by Quantum Computing}% Force line breaks with \\
\author{Cristian Micheletti}
\affiliation{SISSA, Via Bonomea 265, I-34136 Trieste, Italy.}
\email{cristian.micheletti@sissa.it}
\author{Philipp Hauke}
\affiliation{Physics Department of Trento University and INO-CNR BEC Center, Via Sommarive 14, I-38123 Povo (Trento), Italy.}
\author{Pietro Faccioli}
\affiliation{Physics Department of Trento University and INFN-TIFPA, Via Sommarive 14, I-38123 Povo (Trento),   Italy.}
\email{pietro.faccioli@unitn.it}

\begin{abstract}
Sampling  equilibrium ensembles of dense polymer mixtures is a paradigmatically hard problem in computational physics, even in lattice-based models. 
Here, we develop a formalism based on interacting binary tensors that allows for tackling  this problem using quantum annealing machines. 
Our approach is general in that properties such as self-avoidance, branching, and looping  can all be specified in terms of quadratic interactions of the tensors. Microstates realizations of different lattice polymer ensembles are then seamlessly generated by solving suitable discrete energy-minimization problems. This approach enables us to capitalize on the strengths of quantum annealing machines, as we demonstrate by sampling polymer mixtures from low to high densities, using the D-Wave quantum computer. Our systematic approach offers a promising avenue to harness the rapid development of quantum computers for sampling discrete models of filamentous soft-matter systems.
\end{abstract}
\maketitle

\emph{Introduction:} Despite exciting recent progress \cite{Arute2019,Zhong2020}, present-day quantum computers cannot yet outperform classical (super)computers in solving challenging physics problems of general interest. However, considering the current growth rate of their performance, it is highly timely to design new algorithmic paradigms for tackling  problems that have proved hard for classical computing, and thus understand the implied advantages and disadvantages.
With its wealth of problems that are inherently hard computationally, classical statistical mechanics offers an ideal avenue for such endeavors. Yet, while the potential of quantum computers has been extensively explored for quantum many-body systems 
\cite{Hempel2018,Cao2019,Genin2019,Outeiral2020,McArdle2020,QNP1,Zohar2015,Dalmonte2016,QQCD,Hauke2011d,Cirac2012},
thus far there have been very few classical statistical-mechanics applications, mostly in biophysics contexts~\cite{hauke2021dominant,QMD1,QMD2,QMD3,robert2021resource}.

In this work, we discuss the use of quantum annealing machines~\cite{Das2008,Albash2018,Venegas-Andraca2018,Hauke2020} to tackle a paradigmatic statistical mechanics problem, namely sampling the equilibrium ensemble of self-avoiding walks and rings, from dilute to concentrated solutions.

\begin{figure}[t!]
\begin{center}
	\includegraphics[width=0.99\columnwidth]{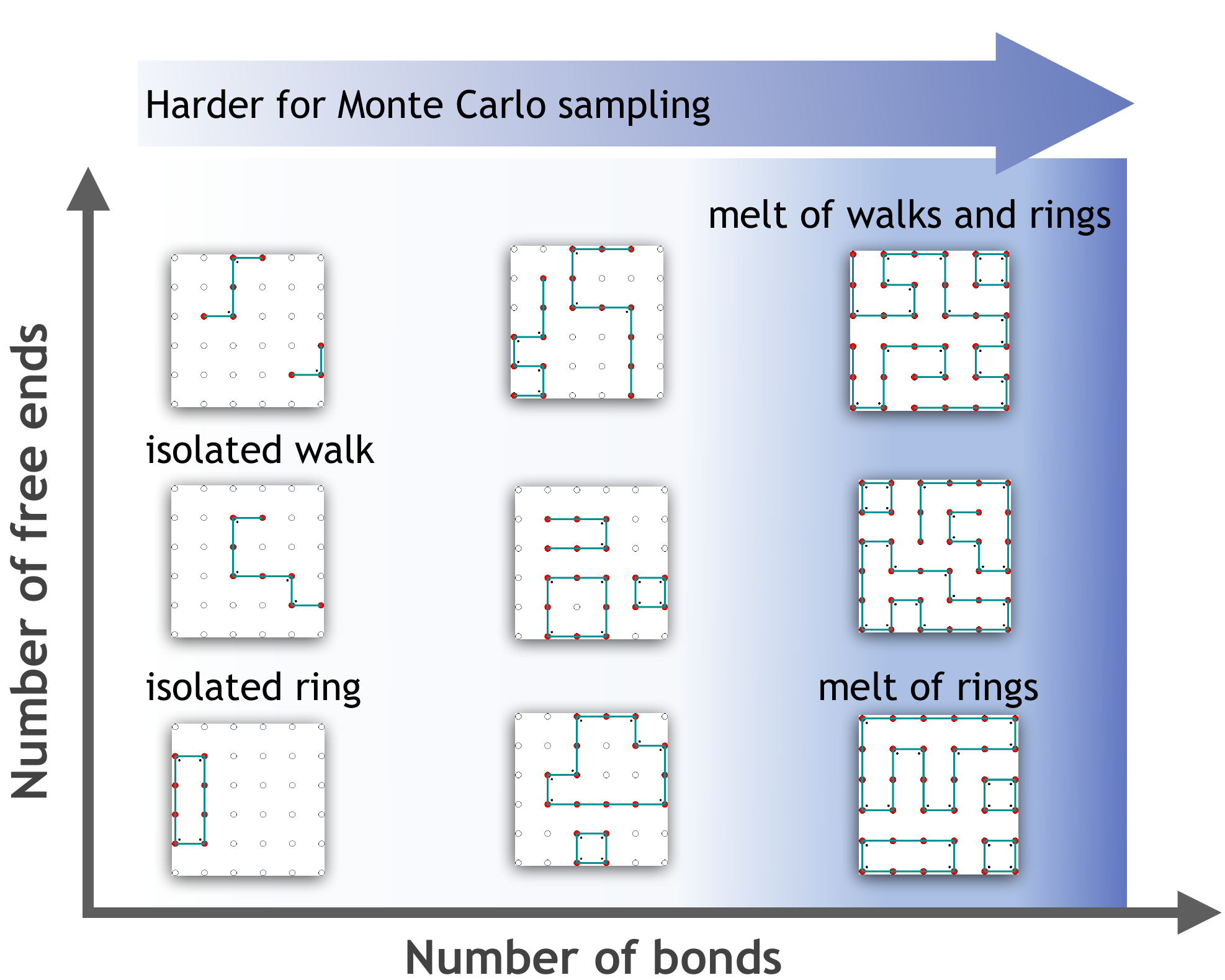}
	\caption{The QUBO formulation of the sampling problem allows for generating mixtures of polymers with discrete degrees of freedom (here schematised for a square lattice embedding) at fixed number of bonds and free ends. Even in the high density limit, which is the most challenging for traditional stochastic sampling schemes, quantum annealing machines maintain their QUBO-solving performance as lattice size increases.}    
\label{fig:1}
\end{center}
\end{figure}

Generating configurations of self-avoiding polymers is an algorithmic challenge that has accompanied computational physics, and contributed to its growth, since its early days. The gist of the challenge is best illustrated for lattice embeddings of self-avoiding walks. As their chain length increases, such paths rapidly become a negligible fraction of all possible walks, thus making it impractical to sample them by discarding \emph{a posteriori} self-crossing conformations from a collection of random paths.
The efforts that have been spent over decades to overcome this attrition problem have given rise to powerful general concepts and methods, from Monte Carlo (MC) with thermodynamic re-weighting~\cite{rosenbluth1955monte} to  multiple Markov chains~\cite{tesi1996monte}.

Elegant methods and strategies are now available to sample self-avoiding walks \cite{madras1988pivot,clisby2010accurate} even of considerable length~\cite{grassberger1997pruned} and enumerate them in bulk or in compact phases \cite{conway1996square,jacobsen2008unbiased}. Nevertheless, efficient sampling of dense solutions or melts of self-avoiding polymers remains a major challenge for both MC and molecular dynamics simulations, because topological constraints create exceedingly long auto-correlation times.

To develop a general framework for polymer sampling on quantum annealers, we introduce a quadratic unconstrained binary optimization (QUBO) problem, where the Hamiltonian is chosen in such a way that its degenerate classical minima  are in bi-univocal correspondence with polymer configurations on a lattice. 
Independent realizations of polymer mixtures at any specified density can be obtained by repeated numerical minimization of the energy function.

In traditional polymer sampling strategies, the length and number of chains are set in the initial state and preserved during the subsequent stochastic evolution of the system. Instead, our QUBO model constrains the total number of monomers $N$ and the number of bonds $L$ in the system or, equivalently, the density (i.e.\ lattice filling fraction) and number of free chain ends in the mixture (see Fig.~\ref{fig:1}). The total number of chains and their lengths can instead fluctuate around their ensemble averages.
Our approach can be seamlessly used to sample different statistical ensembles by introducing or removing energy penalties, e.g.\ for branching or minimum size of admissible loops.

\begin{figure}[t!]
\begin{center}
	\includegraphics[width=0.49\textwidth]{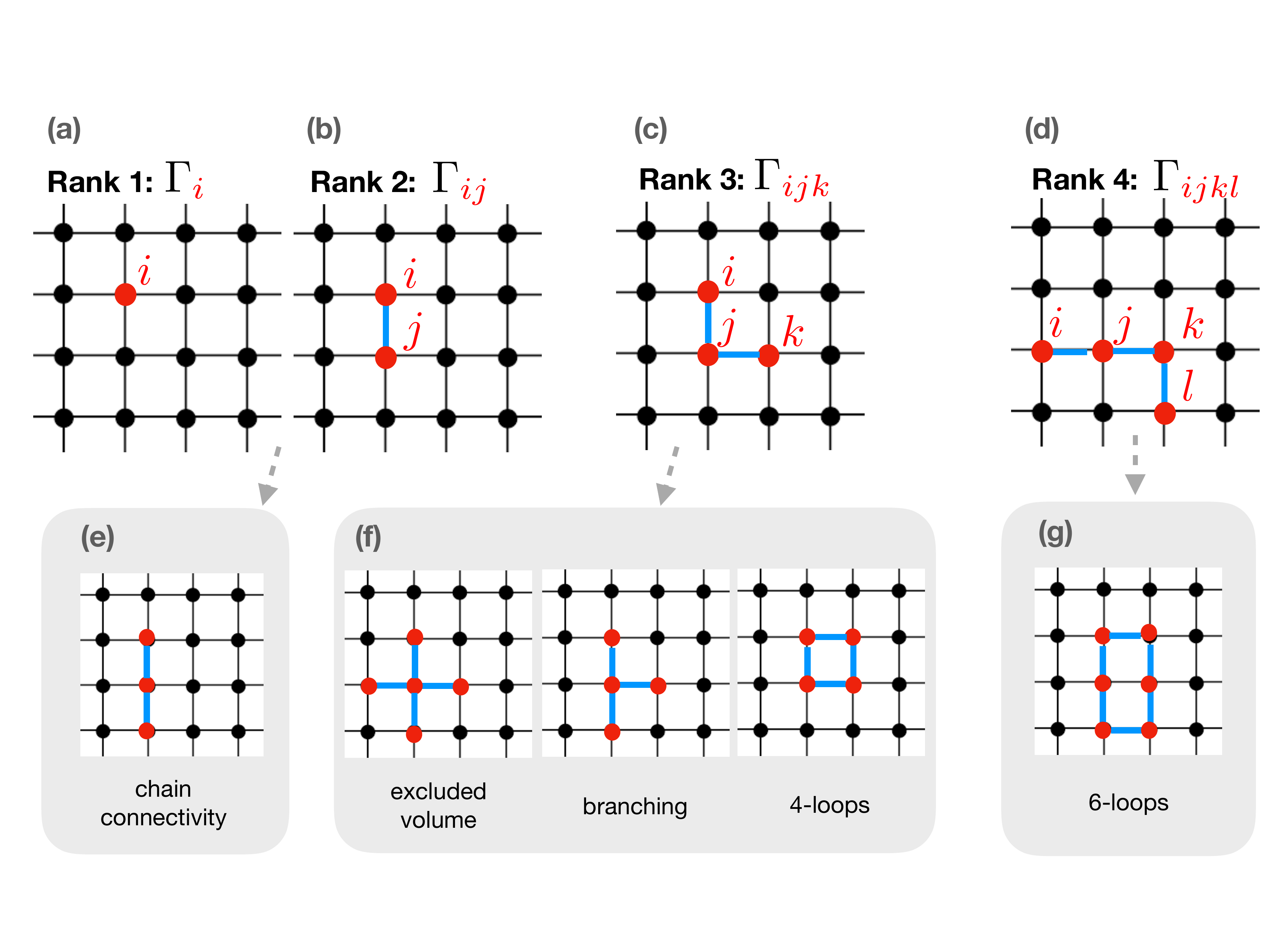}
	\caption{Top row:  representation of the binary tensors, exemplified for a square lattice.  Bottom row: specific features or properties of the polymer chains can be specified via quadratic interactions involving tensors of appropriate rank.}
\label{fig:2}
\end{center}
\end{figure}

\emph{QUBO Hamiltonian:} Our QUBO model is defined in terms of binary tensors (BTs) of different ranks, with the tensor indices running over the sites in the embedding lattice. In a quantum annealing machine, a physical qubit is assigned to each entry of the BTs.  
In the logical problem,  the element $\Gamma_i$ of the rank-1 BT is associated to the $i-$th site (Fig.~\ref{fig:2}a). The elements $\Gamma_{ij}$ of the rank-2 BT correspond to non-oriented bonds between neighboring sites $i$ and $j$ (Fig.~\ref{fig:2}b). Analogously, the entries of the rank-3 BT, $\Gamma_{ijk}$, and the higher order ones are defined in terms of triplets or multiplets of distinct sites that yield connected paths when neighbouring sites are bridged, see e.g.\ Fig.~\ref{fig:2}c-d.
Each tensor entry is a binary variable that assumes the values 1 or 0 if the corresponding site, bond, or $n$-plet is active (occupied) or inactive (empty), respectively.

We show below that by systematically  adding suitable interactions between the tensors  it is possible to enforce the desired physical constraints in the corresponding polymer ensemble, from the total chain lengths and chain connectivity, to self-avoidance and no branching. 
The Hamiltonian formulation provided hereafter is entirely general, though for simplicity we shall discuss it for square lattice embeddings. 
Importantly, it is formulated to involve at most quadratic functions of the BTs, as required by current quantum annealing machines \cite{Hauke2020}.

The simplest polymer ensemble in this framework is generated by enforcing only the constraints of total number of monomers and bonds, and chain connectivity. This requires only a quadratic Hamiltonian of rank-1 and  rank-2 tensors,  
$
H_0 = V_{\textrm{mon}} + V_{\textrm{bond}} + V_{2}\,
$, where
\begin{eqnarray}
V_{\textrm{mon}} &=& A_{\textrm{mon}} \left(\sum_i \Gamma_i - N\right)^2\ , \label{eqn:v1}\\
V_{\textrm{bond}} &=& \frac{A_\textrm{bond}}{2} \left({\sum}'_{i, j} \Gamma_{ij} -L\right)^2\ , \label{eqn:v2}\\
V_{2} &=& \frac{A_{2}}{2} {\sum}'_{i, j} \Gamma_{ij} (1-\Gamma_i). \label{eqn:v3}
\end{eqnarray}
\noindent  Here and below, all coupling constants are assumed to be positive. The prime in ${\sum}'$ denotes summation over distinct running indices. 
Upon energy minimization, interaction terms (\ref{eqn:v1}) and (\ref{eqn:v2}) select polymer mixtures covering $L$ bonds and $N$ monomers (sites) in total. 
Term (\ref{eqn:v3}) instead enforces consistency of the chain connectivity,  penalizing cases where an active  bond is flanked by at least one inactive site (see Fig.\ref{fig:2}e). 
 \begin{figure*}[t!]
\begin{center}
	\includegraphics[width=0.8\textwidth]{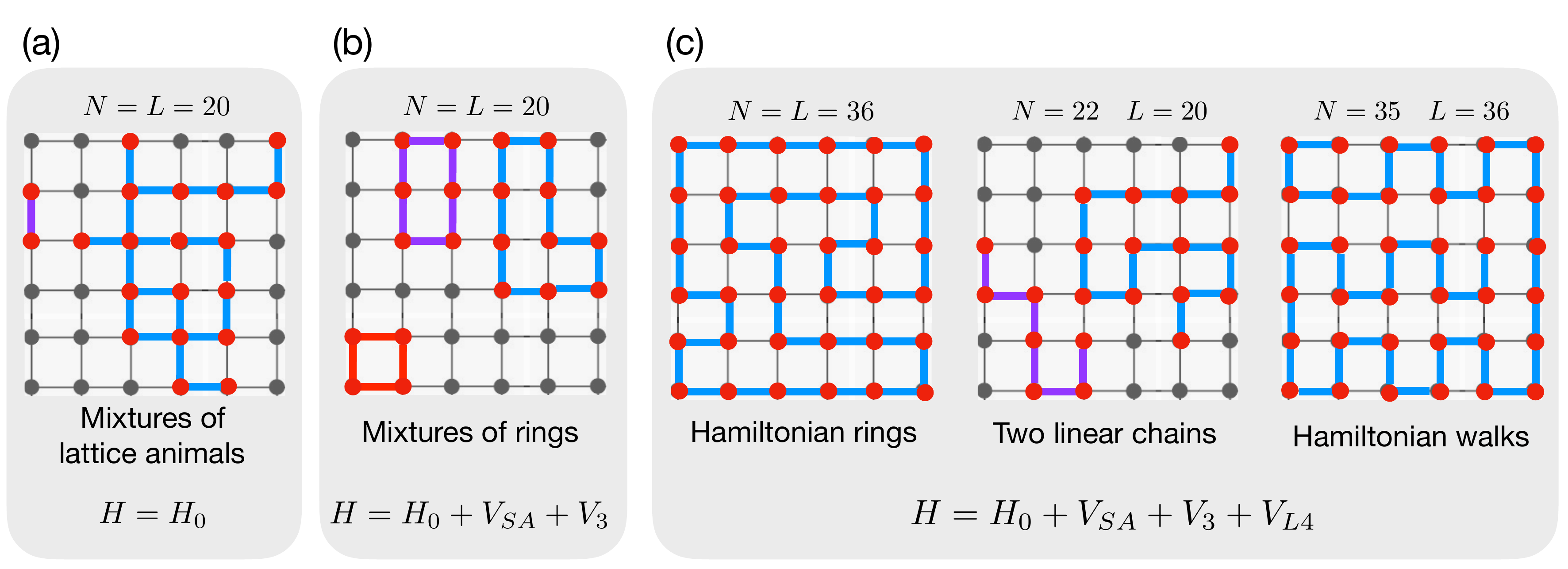}
	\caption{Examples of the different types of polymer mixtures obtained by minimizing the indicated QUBO Hamiltonians for a 6$\times$6 lattice. Active sites and bonds are highlighted in color. Here, and for the results of subsequent figures, we set $A_{\textrm{site}}=1$, $A_{\textrm{bond}}=A_{2}=2$, $A_{\mathrm{SA}}=5!$, $A_{3}=3!$, $A_{L4}=4!$. }
    \label{fig:3}
\end{center}
\end{figure*}
As shown in Fig.~\ref{fig:3}a,  minimizing $H_0$ yields mixtures of lattice animals, a key class of polymers \cite{parisi1981critical,rosa2014ring,everaers2017flory} relevant in percolation theory, too.  

The degenerate ground-state manifold of $H_0$ thus includes configurations that are not self-avoiding and have branches.
Self-intersections and branching are ruled out by complementing $H_0$ with two interaction terms involving the rank-3 tensor.  
The first term is 
\begin{equation}
V_{\mathrm{SA}}= \frac{A_{\mathrm{SA}}}{5!}{\sum}'_{i,j,k, l, m}
\Gamma_{ijk} \Gamma_{ljm},\label{eq:Vsaw}
\end{equation}
\noindent with the proviso that the middle index, $j$, refers to the interior site of the corresponding lattice trimer, see Fig.~\ref{fig:2}c. The term in Eq.~(\ref{eq:Vsaw}) penalizes cases where two active trimers share the middle site, a condition realized at crossing and branching points, see e.g.\ Fig.~\ref{fig:2}e.
In addition, in analogy with Eq.~(\ref{eqn:v3}), the consistency of active elements of the rank-3 tensor with those of lower rank, must be enforced via the following term
\begin{equation}
V_{3}=\frac{A_{3}}{3!} {\sum}'_{i, j, k} \left[3 \Gamma_{ijk} + \Gamma_{ij}\Gamma_{jk}    -2 \Gamma_{ijk} (\Gamma_{ij}+ \Gamma_{jk})\right]. \label{eq:Vgamma}\\
\end{equation}

Minimizing the Hamiltonian $H = H_0 + V_{\mathrm{SA}} + V_{3}$ yields the desired mixtures of self-avoiding chains with no branching.
Mixtures exclusively involving ring polymers are obtained by setting $N=L$.
Typical configurations at different filling fractions of a $6\times6$ square lattice are shown in Fig.~\ref{fig:3}b-c, see SM for larger lattices.

Our QUBO Hamiltonian is particularly suitable for being minimized by quantum annealers, as we discuss below with a direct application on the D-Wave quantum computer. We further note that the resource requirements of the QUBO model are set solely by the number of tensor elements, and not by the number of active bonds or sites, and thus dilute and dense mixtures and melts are dealt with on equal footing. In particular, in straightforward implementations of the Hamiltonian $H_0+V_{SA}+V_3$ for a square lattice with  $n$ sites per side,  the number of required qubits is $7n^2 -10n+4$ (see SM). Such implementation take advantage of the fact that, for better resource efficiency, only non-collinear trimers (i.e.~two bonds meeting at an angle) need to be included in interactions involving the rank-3 BT.

\begin{figure}[b!]
\begin{center}
	\includegraphics[width=0.5\textwidth]{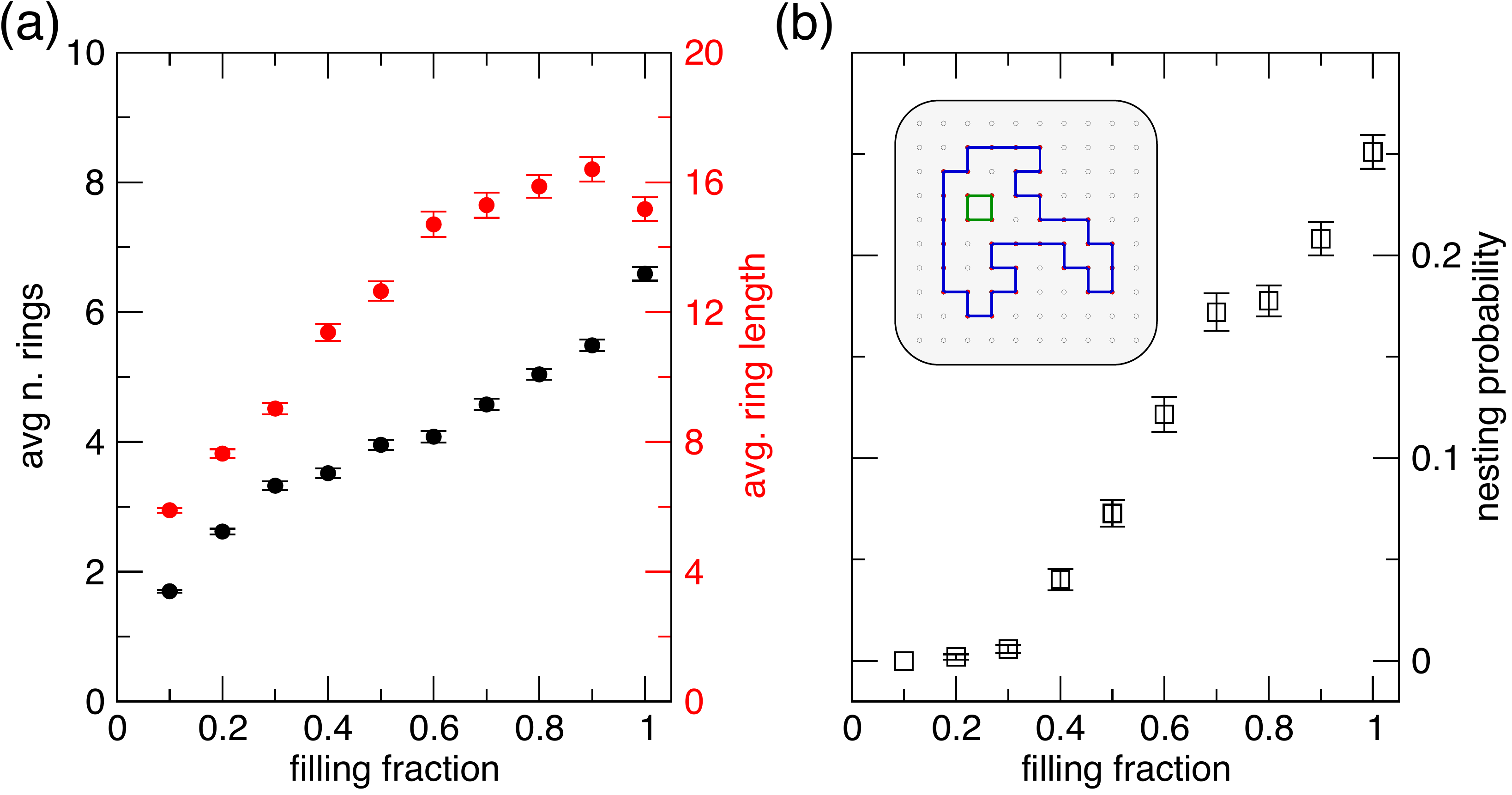}
	\caption{(a) Average number of rings and their length in ring mixtures at varying filling fraction of a 10$\times$10 lattice. (b) Probability that any given ring is involved in one or more nestings. At each filling density, at least 400 samples were obtained  by using a classical simulated annealer to minimize $H=H_0+V_{\mathrm{SA}}+V_3$ using the same interaction coefficients given in Fig.~\ref{fig:3} and with $L=N$.  The inset shows a nesting involving two rings at 0.4 filling fraction.}
\label{fig:4}
\end{center}
\end{figure}

\emph{Mixtures of rings:} An application of our  QUBO method to sample ring mixtures on a $10\times10$ lattice is given in Fig.~\ref{fig:4}. 
As the fraction of occupied lattice sites grows, the average number of rings in the mixtures and their lengths increases too, following the balance of two entropic terms. The first regards the number of distinct ring shapes, which grows approximately exponentially with the ring length, while the second is the roto-translational entropy. As density increases, it becomes advantageous to have more rings in the mixture. Indeed, the gain in roto-translational entropy of many but small rings dominates over the loss in conformational entropy as compared to larger but fewer rings.

The sampled mixtures include instances where one or more rings are fully contained inside larger ones. Figure~\ref{fig:4}b, shows that the probability for any ring to be involved in such nestings increases steadily with density, even at filling fractions larger than 0.6, where we find the variation of average ring length to be limited. 
This result could be relevant in more realistic contexts, such as adsorbates of circular DNA rings \cite{witz2011conformation} or solutions of uncatenated rings \cite{michieletto2014threading,smrek2016minimal,smrek2019threading}, for which systematic studies for the incidence of nestings and threadings (their off-plane generalization) at varying density are not yet available. These considerations show that, even in purposely minimalistic contexts, the QUBO sampling method can illuminate current problems, and thus open promising avenues for larger scale applications.

\emph{Mixtures with linear chains:}
Mixtures that include linear chains can be straightforwardly generated, too.
The number of linear chains, $n_l$, can be specified by assigning $N=L+n_l$.
The number of rings in such mixtures, as well as the contour lengths of the linear and circular chains are, again, controlled by same entropic effects discussed earlier. In fact, low filling fractions will yield configurations that mostly consist of $n_l$ linear chains only, see Fig.~\ref{fig:1}. 

Thus, at low density, setting $N=L+1$ will mostly return individual linear chains of length $L$. Likewise, with $N=L$ one mostly obtains individual rings. 
Our QUBO Hamiltonian approach can be extended to generate mostly or exclusively single rings and walks, and mixtures with linear chains only, also in non-dilute conditions.
This can be accomplished by systematically introducing energy penalties for rings with up to  $L-1$ bonds, at the cost of increasing the number of necessary qubits.

It is convenient to illustrate this procedure by considering the suppression of the shortest possible rings consisting of 4 bonds, 4-loops in brief. 
In our approach, this is done by introducing the following interaction that involves the rank-3 BT:
\begin{eqnarray}
V_{L_4}= \frac{A_{L_4}}{4!} {\sum}'_{i, j, k, l} 
\Gamma_{ijk} ~\Gamma_{kli}\ .
\end{eqnarray}
\noindent  This term penalizes instances where two non-collinear active trimers share the two endpoints (see last case of Fig.~\ref{fig:2}f).  Suppressing such loops boosts the occurrence of single-component configurations, even in the worst case scenario of Hamiltonian walks, as shown in the SM.

The elimination procedure can be generalised to loops of any size by systematically introducing interactions with higher-rank BTs, e.g.\ rank-4 tensors for 6-loops, see Fig. \ref{fig:2}g.  This can be accomplished with the recursive scheme reported in the SM (see Fig. \ref{fig:2}g.

 \begin{figure}[t!]
\begin{center}
	\includegraphics[width=7 cm]{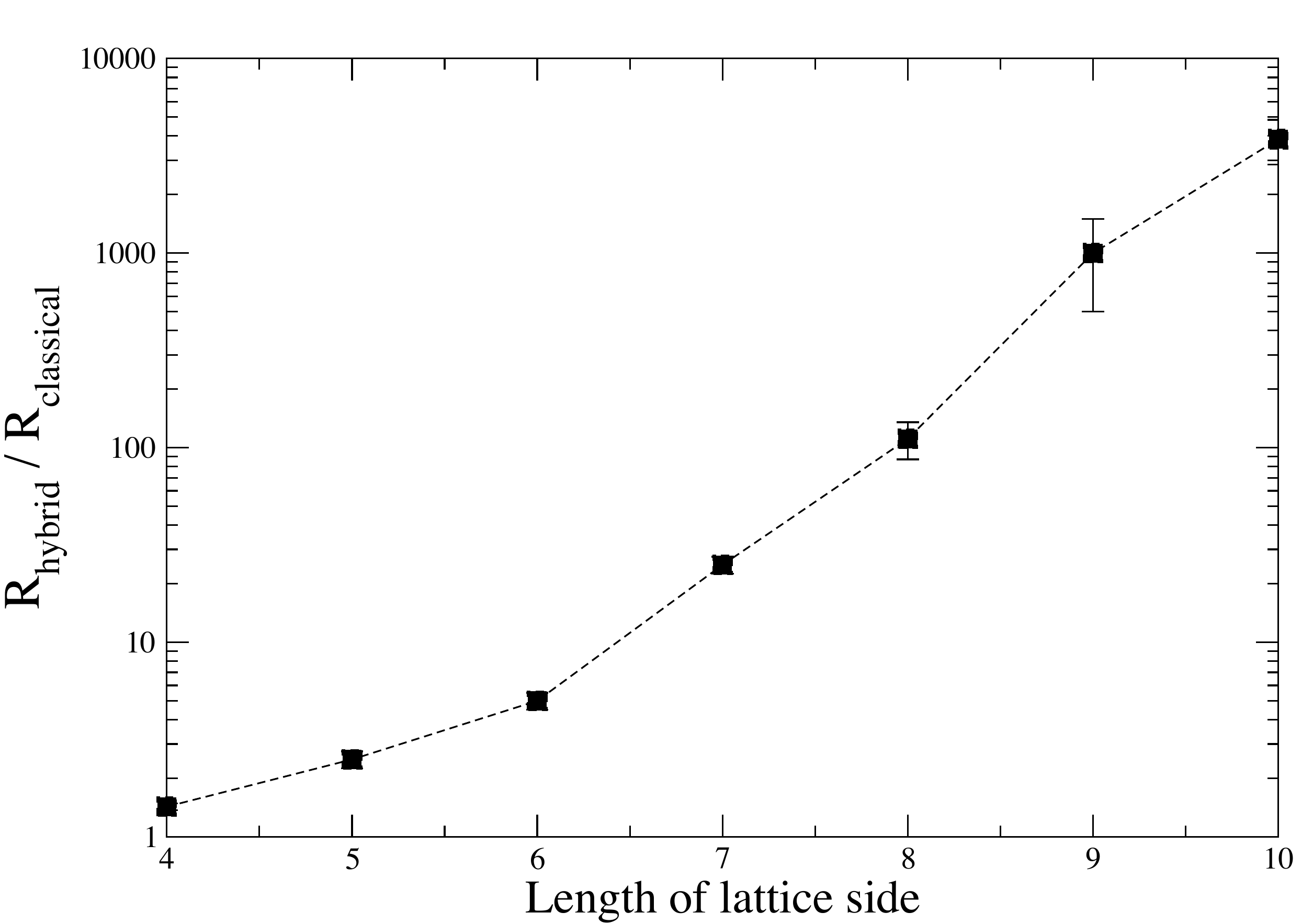}
	\caption{Relative success rate of minimizing the QUBO Hamiltonian $H=H_0 + V_{SA} + V_3$ using the D-Wave hybrid (classical/quantum) and fully-classical simulated annealer. Data are for square lattices of different size at maximum filling fraction. The success rate, $R$, is computed as the fraction of annealing cycles that successfully reach the degenerate ground state manifold.}
\label{fig:5}
\end{center}
\end{figure}

\emph{Computational performance of quantum annealing:} We conclude by discussing the feasibility of minimizing our QUBO Hamiltonian with present-day quantum annealers and comparing it with a standard simulated annealing approach. 
For calculations on a quantum annealer, each entry of the BTs is assigned to a qubit. 
In the standard setting~\cite{Das2008,Albash2018,Venegas-Andraca2018,Hauke2020}, the qubits are initialized in the ground state of an easily solvable Hamiltonian $H_\mathrm{in}$ that does not commute with $H$. Then, the system's Hamiltonian is gradually changed with time, $t$, according to a given schedule $H(t) = a(t) H_\mathrm{in} + b(t) H$. The functions $a(t)$ and $b(t)$ are chosen so that $a(0)=1$ and $b(0)=0$, while at the end of the protocol, $t=t_\mathrm{sweep}$, one has $a(t_\mathrm{sweep})=0$ and $b(t_\mathrm{sweep})=1$. The adiabatic theorem ensures that if the sweep $H_\mathrm{in} \to H$ is sufficiently slow, the final state at the end of the protocol is the classical ground-state of $H$, i.e.\ the solution of our QUBO problem. In classical calculations based on standard simulated annealing, instead, changes in the binary variables are accepted with the Metropolis criterion and the system's nominal temperature is gradually lowered.

For an equal footing comparison,
we used the energy-minimization algorithms implemented in Leap~\cite{Leap}, the system operating the D-Wave supercomputer, which offers a fully-classical simulated annealing protocol and several quantum ones. Here, we focused on the Hamiltonian $H_0+V_{SA}+V_3$ and we used the hybrid quantum/classical protocol available in Leap, with default settings for $H_\mathrm{in}$  and scheduling.

We compared the success rate per annealing cycle for reaching the degenerate ground-state manifold with the hybrid and the fully-classical simulated annealing  protocol,  using different lattice sizes and maximum filling. 
We found that the success rate of the classical simulated annealer decayed exponentially with increasing lattice size. In striking contrast, the hybrid algorithm always converged to the correct energy minimum in a single cycle, i.e.~with 100\% success rate. Therefore, the relative performance advantage of the quantum-based minimization protocol as compared to classical simulated annealing grows exponentially with lattice size, see Fig.~\ref{fig:5}.
We also found that the D-Wave quantum computing time per annealing cycle was about equal to 130 ms, irrespective of the lattice sizes.

\emph{Conclusions and outlook:} Applications of quantum computing to polymer 
physics have been few and, to our knowledge, 
mostly directed at lattice models of proteins for identifying the lowest-energy state of a given sequence~\cite{QMD1,QMD2,QMD3,robert2021resource}. In these elegant studies, however,  the conformational space was not sampled,  but  exhaustively enumerated in advance, a feat that becomes formidable as chain length increases.

In contrast, in this work, we presented a 
 systematic theoretical framework to sample equilibrium ensembles of lattice polymer mixtures at any density.

Different physical properties, such as self-avoidance or branching, can be selected for by simply  switching on or off suitable  sets of interactions in a QUBO Hamiltonian. In specific applications, our general approach may be refined in order to reduce the number of required qubits, as illustrated in the SM. 
Our QUBO formulation is particularly suitable for the quantum annealing approach even in present-day devices, as we showed with direct applications on the D-Wave quantum computer. We thus expect that the accelerating pace of quantum computing innovation will make it possible to extend our approach to more refined and larger scale models, and thus tackle challenging systems that are of current interest.  These include polymer melts, where topological constraints and threadings, which generalize the nestings of Fig.~\ref{fig:4}, can hinder the system evolution and hence sampling within traditional simulation schemes.

In addition, while in this first study we focused on fully-flexible homopolymers, an interesting development would be to extend our approach to semiflexible chains, and to heteropolymers. We envisage that this could be conveniently done in a dual classical/quantum approach, i.e.\ by combining QUBO  with thermodynamic re-weighting strategies performed on classical computers. 

Finally, we note that sampling applications with quantum annealers have been discussed before in machine learning and other contexts \cite{Adachi2015,Szoke2016,Benedetti2017,Sieberer2018,Vinci2019,Koenz2019,Yamamoto2020,Kumar2020}, and thus we hope that our work can also inspire follow-up investigations connecting the present scheme and such studies. 

\emph{Acknowledgments:} We thank D-Wave for granting free access to their quantum annealing machine and A.~Rosa, G.~Mazzola and G.~Mattiotti for useful discussions. 
P.H.~acknowledges support by Provincia Autonoma di Trento, the Bundesministerium für Wirtschaft und Energie through the project "EnerQuant" (ProjectID 03EI1025C), and
the ERC Starting Grant StrEnQTh (Project-ID 804305).  P.H.~and P.F.~participate in the Q@TN|Quantum Science and Technology initiative in Trento. 
%\bibliographystyle{}

%\bibliographystyle{abbrv}
%\bibliography{bibliography}

\newpage\mbox{}
\newpage
\renewcommand{\theequation}{S\arabic{equation}}%labels equations with Eq. (S1) etc. 
\setcounter{equation}{0}%sets counter for the equations to 1, so that start with (S1)
\renewcommand{\thefigure}{S\arabic{figure}}%the same for the figures
\setcounter{figure}{0}

\noindent{\Large \bf Supplementary Material}\\

\vspace{0,5cm}

In this Supplemental Material, we present: (i) a counting of the number of qubits required to implement the model on square lattices, (ii) algorithmic details for calculating the number of rings involved in nestings, (iii) numerical results for ensembles obtained with energy penalties for 4-loops, (iv) a general algorithmic prescription for removing loops of arbitrary length by introducing additional binary tensors, and (v) a modification of our QUBO Hamiltonian to impose self-avoidance and absence of branches using fewer qubits than the general algorithm presented in the main text.

\section{Number of qubits used in square lattice implementations of the QUBO model.}

The table below shows the the counting of qubits used in straightforward implementations our QUBO model for a square lattice with $n$ sites per side. For each value of $n$ the Table provides the number of binary variables we used to encode sites ($n_s$), bonds ($n_b$), non-collinear trimers ($n_t$). Their sum is provided in the rightmost column and corresponds to the total number of qubits used to minimize the QUBO Hamiltonians $H=H_0+V_{\mathrm{SA}}+V_{3}$ and $H=H_0+V_{\mathrm{SA}}+V_3+V_{\mathrm{L4}}$ .

\begin{center}
\begin{tabular}{ || c | c | c | c || c || } 
 \hline
 Lattice side & $n_s$ & $n_b$ & $n_t$ & $n_{\rm qubits}$ \\ \hline \hline
 2 & 4  &  4 &  4 & 12 \\
 3 & 9  & 12 & 16 & 37\\
 4 & 16 & 24 & 36 & 76\\
 5 & 25 & 40 & 64 & 129\\
 6 & 36 & 60 & 100 & 196\\
  $\vdots$ & & $\vdots$ & & $\vdots$ \\ 
 $n$ & $n^2$ & $2n(n-1)$ & $4 (n-1)^2$ & $7n^2 -10n +4$\\
 \hline
\end{tabular}
\end{center}

\section{Assessing incidence of nested rings}

\begin{figure}[h!]
\begin{center}
	\includegraphics[width=8 cm]{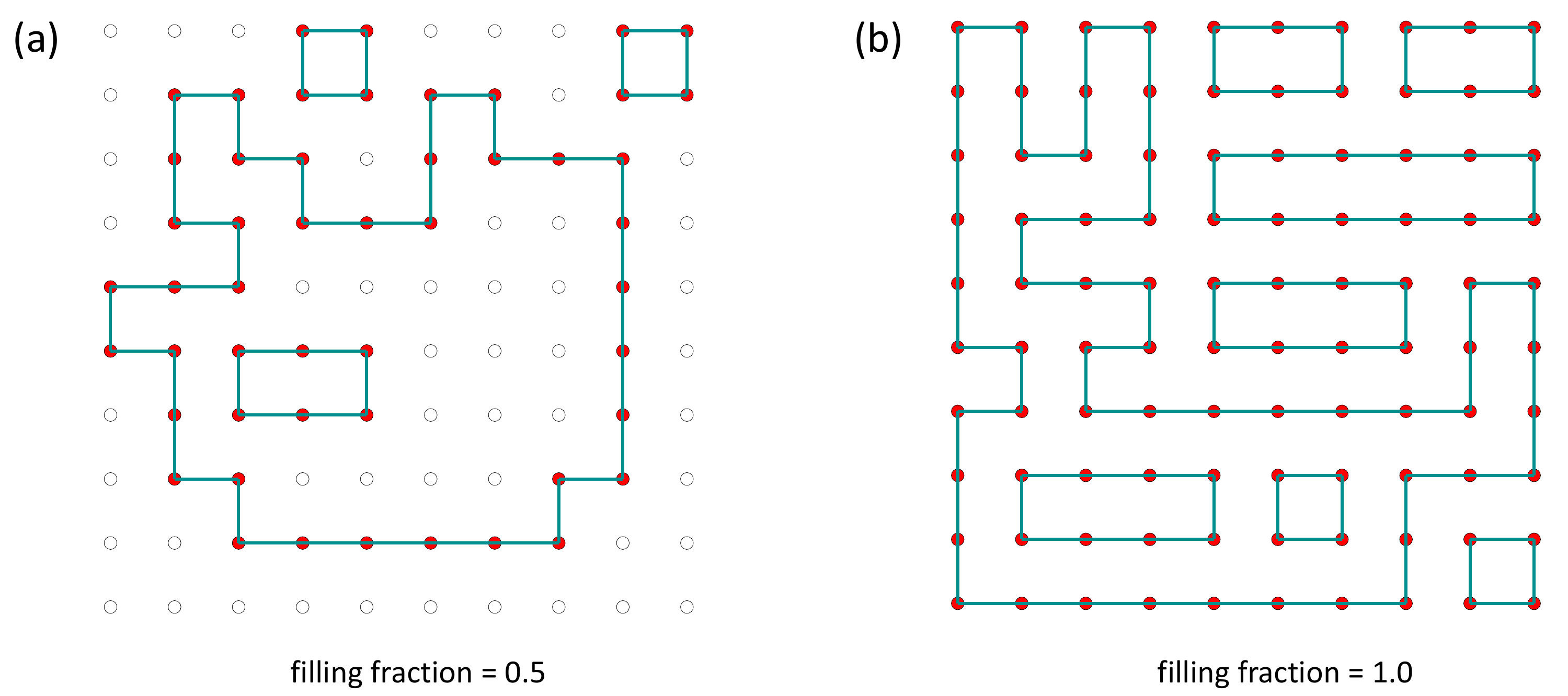}
	\caption{Examples of configurations with nested rings observed in  mixtures at the indicated filling fractions of a 10$\times$10 lattice.
	Active bonds and sites (monomers) are highlighted in color.  The number of rings involved in nestings is 2 and 3 for panels (a) and (b), respectively. Mixtures were sampled by minimizing $H=H_0+V_{\mathrm{SA}}+V_{3}$ with $L=N$ and with the same interaction coefficients as reported in Fig.~3 of the main text.}
%	\label{fig:concept}
\label{fig:S1}
\end{center}
\end{figure}

Samples of ring mixtures obtained by minimizing the QUBO Hamiltonian presented in the main text naturally include instances of nested rings, as illustrated in Fig.~\ref{fig:S1}.

The counting of such nestings as presented in Fig.~4 of the main text was broken down in the following algorithmic steps. For each configuration, we first computed the adjacency matrix of active bonds and then used it to identify the distinct components, i.e.\ the various rings. Next, for each ordered pair of rings, we established whether the first ring was entirely contained (nested) inside the second. This involved checking whether all monomers of the first ring where inside the polygon defined by the second ring. The point-inside-polygon condition was assessed with the winding number criterion. If the nesting condition was realized, both the first (guest) and second (host) rings were flagged. 
After scanning all ordered ring pairs, each flagged ring contributed to the count of rings involved in nestings.

\section{Removing loops of length 4}

\begin{figure}[t!]
\begin{center}
	\includegraphics[width=7 cm]{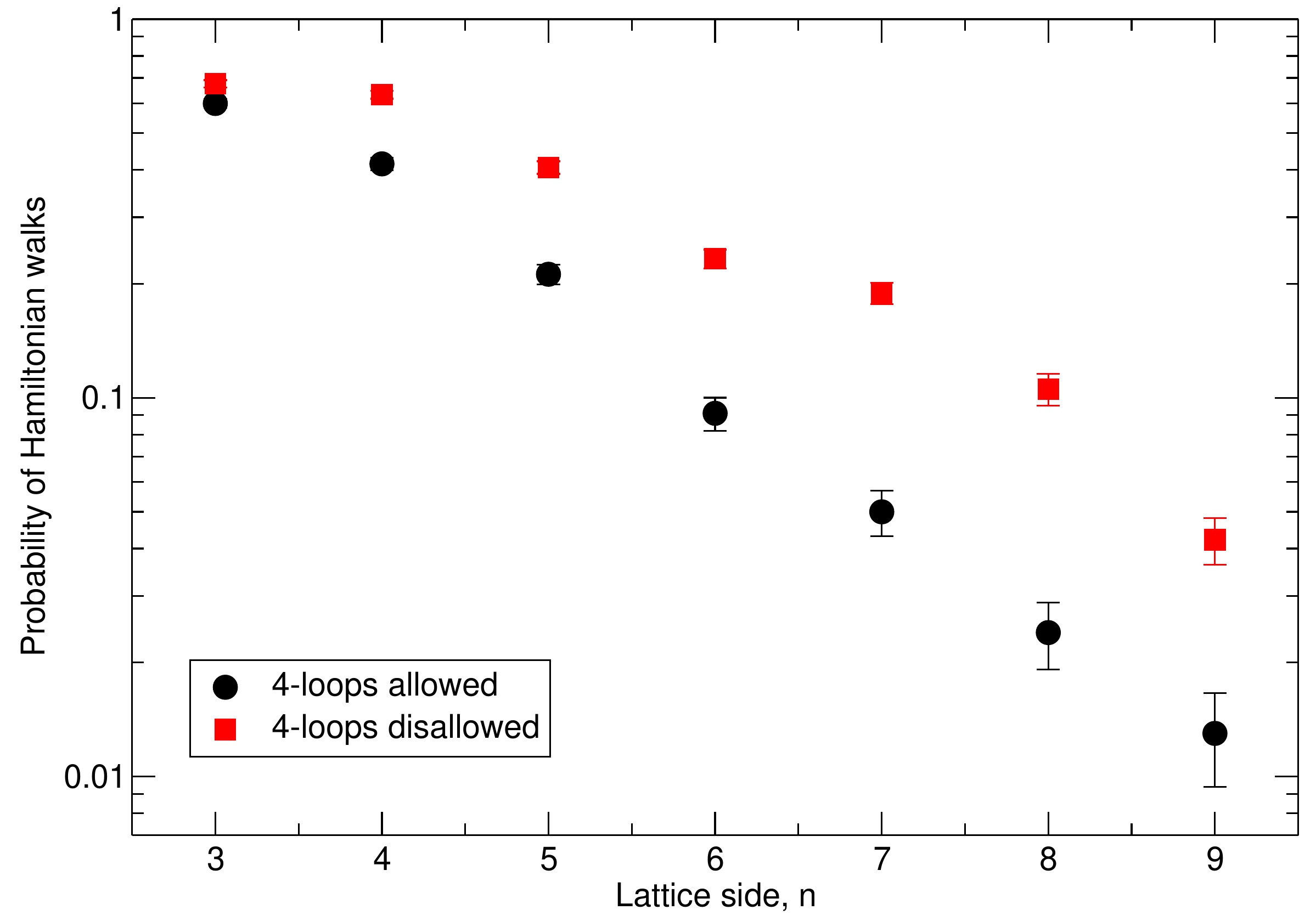}
	\caption{Probability that the QUBO procedure samples a (single) Hamiltonian walk of length $L=n^2-1$ on an $n\times n$ lattice when 4-loops are allowed or disallowed. The corresponding QUBO Hamiltonians were $H=H_0+V_{\mathrm{SA}}+V_{3}$ and $H=H_0+V_{\mathrm{SA}}+V_{3}+V_{L4}$, respectively. Probabilities and estimated errors are computed from samples of at least 1,000 configurations at each value of $n$. The mixtures were sampled by minimizing the QUBO Hamiltonian using the same interaction coefficients given in Fig.~3 of the main text and setting $A_{\mathrm{L4}}=4!$ and $N=L+1$.}
%	\label{fig:concept}
\label{fig:S2}
\end{center}
\end{figure}

Loops of length 4 can be eliminated by adding to the Hamiltonian the energy term of Eq.~(6) in the main text. Suppressing such loops, which have the shortest possible length, already boosts the probability to sample single walks or rings (for $N$ equal to $L+1$ and $L$, respectively). This is illustrated in Fig.~\ref{fig:S2} for the case of Hamiltonian walks on lattices of increasing size.  The case of Hamiltonian (space filling) walks represents a worst case scenario as the average number of different components grows with the lattice filling fraction, see Fig.~4 in the main text.

\section{Removing Loops of Arbitrary Length}
In this section, we illustrate the procedure to eliminate rings of arbitrary length from the ensemble of polymer configurations generated by our QUBO. This formulation generalizes that of removing the 4-loops discussed in the main text [see main-text Eq.~(6)]. 

We begin by explicitly reporting the interaction that removes 6-loops, which involves an interaction term among elements of the rank-4 binary tensor  (BT), $\Gamma_{ijkl}$, 
\begin{eqnarray}
V_{L_6}= \frac{C_{L_6}}{6!} {\sum}'_{i,j,k,l,m,n}
\Gamma_{ijkl} \Gamma_{lmni} \ ,
\label{eq:HL6}
\end{eqnarray}
\noindent complemented by a term imposing the consistency of $\Gamma_{ijkl}$ with lower rank variables, analogously to Eq.~(5) of the main text:
\begin{eqnarray}
V_{4} =\frac{C_{4}}{4!} {\sum}'_{i,j,k,l}
\left[3\Gamma_{ijkl} + \Gamma_{ijk}\Gamma_{jkl}-2 \Gamma_{ijkl}  (\Gamma_{ijk}+ \Gamma_{jkl})\right].\nonumber\\
\label{eq:Vgamma4}
\end{eqnarray}
In these equations and in the following,  the prime in ${\sum}'$ denotes  a summation over distinct running indices.

The same procedure can be used to remove loops of arbitrary length. Namely, the interaction that removes the loops of order $2k$ is defined as a quadratic interaction between elements of BTs of rank $k+1$ $\Gamma_{i_1, \ldots, i_{k+1}}$, 
\begin{eqnarray}
V_{L_{2k}} = \frac{A_{L_{2k}}}{(2k)!}{\sum}'_{i_1,\ldots, i_{2k}}
\Gamma_{i_1 i_2 \ldots i_{k+1}}\Gamma_{i_{k+1} i_{k+2} \ldots i_{2k} i_1}.
\end{eqnarray}
Because the first and last indices coincide, the site indices involved in the interaction form a $2k$-loop, which thus gets penalized.  
The consistency relation between the rank $k+1$ BT, $\Gamma_{i_1, \ldots, i_{k+1}}$, and the lower-rank BT, $\Gamma_{i_1, \ldots, i_{k}}$, is imposed by the interaction 
\begin{eqnarray}
&&V_{k+1} =  \frac{A_{\Gamma_{k+1}}}{(k+1)!}{\sum}'_{i_1, \ldots, i_{k+1}} \left[3\Gamma_{i_1, \ldots, i_{k+1}}+ \right.\nonumber\\
&&\left. \Gamma_{i_1, \ldots, i_{k}}\Gamma_{i_2, \ldots, i_{k+1}} -2 \Gamma_{i_1, \ldots, i_{k+1}}(\Gamma_{i_1,\ldots,i_{k}}+ \Gamma_{i_2, \ldots,i_{k+1}})\right].\nonumber\\
%$$\right.\nonumber\\
%&& \left.\right] \nonumber\\
\end{eqnarray}

\section{Self-avoiding and branchcless walks and loops using fewer qubits}
In the main text, we presented a systematic way of generating mixtures of polymers with desired properties. That formulation involved binary tensors of increasing ranks, corresponding to lattice sites, bonds, trimers, and so on. For specific scenarios, it is possible to reduce the number of degrees of freedom that need to be realized in the quantum-annealing machine by suitably modifying or specializing this scheme. 

To illustrate this possible resource economization, we consider the mixture of self-avoiding and branchless chains. These can be realized by using two sets of rank-1 tensors:  $\Gamma_i^\mathrm{int}$, which actives  only when the site $i$  belongs to the internal part of a chain and $\Gamma_i^\mathrm{end}$, which is active only if it belongs to an endpoint. These are augmented by the usual rank-2 tensors $\Gamma_{ij}$ that encode the presence of a bond between sites $i$ and $j$. 
The following energy penalties then give the desired ground-state manifold,  
\begin{eqnarray}
V_{\textrm{int}} &=& A_{\textrm{int}} \left(\sum_i \Gamma_i^\mathrm{int} - N_\mathrm{int}\right)^2\ , \\
V_{\textrm{end}} &=& A_{\textrm{end}} \left(\sum_i \Gamma_i^\mathrm{end} - N_\mathrm{end}\right)^2\ , \\
V_{2,\textrm{int}} &=& \frac{A_{2,\textrm{int}}}{2} \sum_{i} \left(\sum_{j\neq i}\Gamma_{ij} -2\Gamma_i^{\textrm{int}}\right)^2, \\
V_{2,\textrm{end}} &=& \frac{A_{2,\textrm{end}}}{2} \sum_{i} \left(\sum_{j\neq i}\Gamma_{ij} -\Gamma_i^{\textrm{end}}\right)^2.
\end{eqnarray}
The first two constraints set the total number of internal monomers,  $N_\mathrm{int}$, and  endpoints, $N_\mathrm{end}$, of the polymers in the mixture (in main-text notation: $N_\mathrm{end}=2(N-L)=2n_l$ and $N_\mathrm{int}=N-N_\mathrm{end}$). 
The last two terms enforce the consistency of active elements of  rank-1 and rank-2 tensors, as well as the self-avoidance and no-branching constraints (active bonds cannot bridge inactive sites; internal and endpoint monomers are flanked by exactly two and one active bonds, respectively). 

This formulation replaces the use of rank-3 tensors, of which there are 4 per lattice site, by only one additional rank-1 tensor per lattice site. 
Further extensions are also possible. For example, 4-loops can be suppressed by adding four-body interactions among the rank-2 tensors of the form $\Gamma_{12}\Gamma_{23}\Gamma_{34}\Gamma_{41}$. One possibility to deal with these non-quadratic interactions would be introducing additional ancillary qubits \cite{Leib2016,babbush2013resource} for mapping them to an equivalent quadratic Hamiltonian. The alternative would be to rely on a new generation of machines able to realize multi-qubit interactions \cite{Hauke2020}. 
As this example shows, our general QUBO formulation invites to modifications whose most resource economic variant depends on the particular scenario of interest and the available machine.

\bibliography{bibliography}

\end{document}